\documentclass[aip.rsi,amsmath,amssymb,floatfix,superscriptaddress]{revtex4-1}
\usepackage{color}
\usepackage{graphicx}
\usepackage{dcolumn}
\usepackage{bm}
\usepackage{times}
\begin{document}

\title{Absolute cross-section normalization of  
magnetic neutron scattering data}

\author{Guangyong~Xu}

\author{Zhijun~Xu}
\author{J. M. Tranquada}
\affiliation{Condensed Matter Physics and Materials Science
Department, Brookhaven National Laboratory, Upton, New York 11973,
USA}
\date{\today}

\begin{abstract}
We discuss various methods to obtain the resolution volume for
neutron scattering experiments, in order to perform absolute normalization
on inelastic magnetic neutron scattering data. Examples from previous 
experiments are given. We also try to provide clear definitions
of a number of physical quantities which are commonly used to 
describe neutron magnetic scattering results, including the dynamic spin 
correlation function and the imaginary part of the dynamic susceptibility.
Formulas that can be used 
for general purposes are provided and the advantages of the different 
normalization processes are discussed.

\end{abstract}

\maketitle

\section{Introduction}

Magnetic neutron scattering is a powerful technique for studying 
the magnetic structure and spin dynamics of various material systems.
Sometimes obtaining only relative scattering intensities is sufficient. 
Increasingly it is necessary
to convert the magnetic scattering intensities to absolute units,
i.e. in terms of magnet moments ($\mu_B$) or spin (S) per site.
The principle of this normalization process has been described 
previously~\cite{Broholm1,Steinsvoll,Ziebeck}. Here we describe practical
details of the normalization process and discuss potential problems. 
In addition,  closely related yet different physical quantities, 
such as the dynamic spin 
correlation function $S({\bf Q},\omega)$ and the imaginary part of the dynamic 
susceptibility ${\chi}''({\bf Q},\omega)$ are  used to describe 
magnetic neutron scattering intensities. It is 
important to clarify the definition of and conversion between these quantities.
These  are the main purposes of this paper.

The neutron scattering  intensity measured at the detector can be written 
as a convolution of the differential
scattering cross-section $\frac{d^2\sigma}{d\Omega dE}$, which depends on the
sample itself, and the instrument resolution function 
$R({\bf Q_0},E_0,{\bf Q},E)$, which is mainly determined by parameters 
related to the instrument set-up~\cite{Chesser1}: 
\begin{equation}
I({\bf{Q}},E)=\int \frac{d^2\sigma}{d\Omega_0 dE_0}R({\bf{Q_0}},E_0,{\bf Q},E)d{\bf Q_0}dE_0,
\label{eqn:1}
\end{equation}
where ${\bf Q}$ and $E$ are the wave-vector and energy transfer. 
In practice,   it is customary to divide
the detector counts by the monitor counts, the latter being inversely 
proportional to the incident neutron wave-vector $k_i$. Therefore the modified 
neutron scattering intensity becomes:
\begin{equation}
\tilde{I}({\bf Q},E)=k_i\int \frac{d^2\sigma}{d\Omega_0 dE_0}R({\bf Q_0},E_0,{\bf Q},E)d{\bf Q_0}dE_0,
\label{eqn:2}
\end{equation}
This is typically the case for measurements done on triple-axis 
spectrometers~\cite{Shiranebook} (TAS). For measurements done on time-of-flight
spectrometers where $k_i$-independent monitors are sometimes used, there are 
usually data reduction options for users to choose, so that the end 
data are presented in a format where $k_i$ and $k_f$ are properly taken care
of. This subtle difference will be discussed again when we 
describe the resolution volume determination in details  below.

Here we only discuss the case of scattering with unpolarized neutrons.
For coherent magnetic scattering from a system with a single species 
of magnetic atom~\cite{Shiranebook}, 
\begin{equation}
\frac{d^2\sigma}{d\Omega dE} = \frac{N}{\hbar}\frac{k_f}{k_i}p^2e^{-2W}
\sum_{\alpha,\beta}(\delta_{\alpha,\beta}-\tilde{Q_\alpha}\tilde{Q_\beta})S^{\alpha
\beta}({\bf Q},\omega).
\label{eqn:3}
\end{equation}
The parameters used in Eq.~\ref{eqn:3} are explained below:

$N$ is the total number of unit cells, 
$p=(\frac{\gamma r_0}{2})gf({\bf Q})$, and 
$\frac{\gamma r_0}{2}=0.2695\times 10^{-12}$~cm, $f({\bf Q})$ is the magnetic form
factor. $k_f$ and $k_i$ are final and incident neutron wave-vectors. $e^{-2W}$ 
is the Debye-Waller factor. 
$\alpha,\beta$ denote the Cartesian Coordinates $x,y$, and $z$, and 
$\tilde{Q_\alpha}, \tilde{Q_\beta}$ are the projections of the unit wave-vector 
$\tilde{Q}$ on the Cartesian axes.
\begin{equation}
S^{\alpha\beta}({\bf Q},\omega) = \frac{1}{2\pi}\int dt e^{-i\omega t}
\sum_l e^{i{\bf Q}\cdot {\bf r_l}}\langle S_0^\alpha(0)S_l^\beta(t)\rangle,
\label{eqn:sqw}
\end{equation}
is the dynamic spin correlation function, which is typically
what one would eventually want to determine. $S^{\alpha\beta}({\bf Q},\omega)$ 
satisfies a simple sum rule when integrated over a Brillouin zone:
\begin{equation}
\frac{\int_{-\infty}^{+\infty} \int_{BZ}S^{\alpha\beta}({\bf Q},\omega)d{\bf Q}dE}
{\int_{BZ}d{\bf Q}} = \frac{1}{3}S(S+1)\delta_{\alpha\beta}.
\label{eqn:4a}
\end{equation}
It is also related to the imaginary part of the dynamic 
susceptibility $\chi''({\bf Q},\omega)$
via the fluctuation-dissipation theorem~\cite{Marshall}:
\begin{equation}
{\chi^{\alpha\beta}}''({\bf Q},\omega) = g^2\mu_{\rm B}^2 \frac{\pi}{\hbar}\Big(1-e^{-\hbar\omega/k_BT}\Big) {S}^{\alpha\beta}({\bf Q},\omega).
\label{eqn:4}
\end{equation}
Note that the prefactors we have included in this definition are necessary in
order that $\chi''$ is consistent with the bulk susceptibility, as we will show
later.

We then make a necessary approximation to ``decouple'' the instrument
resolution and the scattering response function. 
\begin{equation}
\tilde{I}({\bf Q},E)\approx \frac{N}{\hbar}p^2e^{-2W}\tilde{S}({\bf Q},\omega)k_fR_0({\bf Q},E),
\label{eqn:5}
\end{equation}
where $R_0({\bf Q},E) = \int R({\bf Q_0},E_0,{\bf Q},E)d{\bf Q_0}dE_0$ is 
the resolution volume, which should only be instrument dependent, and will
later be referred to as $R_0$. $\tilde{S}({\bf Q},\omega) = 
\sum_{\alpha,\beta}(\delta_{\alpha,\beta}-\tilde{Q_\alpha}\tilde{Q_\beta})S^{\alpha \beta}
({\bf Q},\omega)$ is the modified dynamic spin correlation function, taking
into account that our scattering measurements are only sensitive to 
spin (fluctuations) along directions perpendicular to the wave-vector
transfers. In the common case of isotropic spin excitations, as in a 
paramagnetic phase, 
$S^{xx}({\bf Q},\omega)=S^{yy}({\bf Q},\omega)
=S^{zz}({\bf Q},\omega)$, 
thus 
\begin{equation}
\tilde{S}({\bf Q},\omega) = 2S^{zz},
\label{eqn:iso}
\end{equation}
The approximation in Eq.~\ref{eqn:5} is acceptable if 
$\tilde{S}({\bf Q},\omega)$ is relatively smooth in the region of interest. 
When the real $\tilde{S}({\bf Q},\omega)$ has sharp features relative to the 
resolution volume, usually the accurate non-distorted 
$\tilde{S}({\bf Q},\omega)$ will 
only be obtainable through a deconvolution process, which is beyond the 
scope of this paper. Nevertheless, this approximation is always valid
if one is interested in the total spectral weight, i.e. ${\bf Q}$ and energy 
integral of $S({\bf Q},\omega)$.  

After putting all the numbers together, one can ultimately write down:
\begin{equation}
\tilde{S}({\bf Q},\omega) = \frac{13.77(\text{barn}^{-1})\tilde{I}({\bf Q},E)}{g^2|f({\bf Q})|^2e^{-2W}Nk_fR_0},
\label{eqn:6}
\end{equation}
where $1$~barn $= 10^{-24}$cm$^2$ is the unit for neutron scattering 
cross-section.
We will see in later discussions that the denominator has the units of 
energy divided by cross-section 
(typically barn$^{-1}\cdot$meV or barn$^{-1}\cdot$eV), 
thus $\tilde{S}({\bf Q},\omega) $ here has a unit 
of meV$^{-1}$ or eV$^{-1}$. Alternatively, if one wants to express 
the results as squared magnetic moment per site, in units of 
$\mu_B^2\cdot$mev$^{-1}$ or $\mu_B^2\cdot$ev$^{-1}$, the appropriate formula is
\begin{equation}
\tilde{M}({\bf Q},\omega)=g^2\mu_B^2\tilde{S}({\bf Q},\omega)=
\frac{13.77(\text{barn}^{-1})\mu_B^2\tilde{I}({\bf Q},E)}{|f({\bf Q})|^2e^{-2W}Nk_fR_0},
\label{eqn:6a}
\end{equation}
which only differs by a factor of $g^2\mu_B^2$ compared to $\tilde{S}({\bf Q},\omega)$.
It is worthwhile to mention that in many literature people use 
expressions such as $I({\bf Q},\omega)$, or
simply $S({\bf Q},\omega)$ when plotting magnetic neutron scattering results. 
In most cases, unless explicitly noted, they actually correspond to either 
$\tilde{S}({\bf Q},\omega)$ or $\tilde{M}({\bf Q},\omega)$ that we discussed
above.

Sometimes it may be preferable to write the magnetic neutron scattering results 
in a form directly related to bulk susceptibility measurements. For such
a purpose, the imaginary part of the dynamic susceptibility 
${\chi^{\alpha\beta}}''({\bf Q},\omega)$ is often used, which can be converted 
from $S^{\alpha\beta}({\bf Q},\omega)$ using Eq.~\ref{eqn:4}.
This quantity is related
to the bulk susceptibility through the Kramers-Kronig transformation:
\begin{equation}
  {\chi^{\alpha,\beta}}'({\bf Q},0) = {1\over\pi}\int d\omega {1\over\omega} {\chi^{\alpha\beta}}''({\bf Q},\omega),
\label{eqn:ext1}
\end{equation} 
For the typical case of paramagnetic scattering from non-interacting spins, 
the bulk susceptibility 
\begin{equation}
 \chi = {g^2\mu_{\rm B}^2\over 3k_BT} S(S+1)={\chi^{\alpha\alpha}}'({\bf 0},0).
\label{eqn:bulk}
\end{equation}
Using Eq.~\ref{eqn:sqw} one can verify that the definition of $\chi^{\alpha\beta}$
in Eq~\ref{eqn:4} satisfies Eq.~\ref{eqn:bulk}. Using 
Eqs.~\ref{eqn:iso}, \ref{eqn:4} and \ref{eqn:6}, we have:
\begin{equation}
\chi''({\bf Q},\omega)={\chi^{zz}}'' = \frac{\pi}{2}\mu_B^2\Big(1-e^{-\hbar\omega/k_BT}\Big)\frac{13.77(\text{barn}^{-1})\tilde{I}({\bf Q},E)}{|f({\bf Q})|^2e^{-2W}Nk_fR_0}.
\label{eqn:6c}
\end{equation}
In the case of anisotropic spin
excitations, to avoid confusion it would make more sense to use the 
different components of ${\chi^{\alpha\beta}}''$ explicitly rather than the general
form ${\chi}''({\bf Q},\omega)$.

The Land\'{e}-$g$ factor, the magnetic form factor $f({\bf Q})$, and the
Debye-Waller factor $e^{-2W}$ are sample dependent and can be estimated.
For example $g \approx 2$, $e^{-2W} \approx 1$ for small ${\bf Q}$ where 
$f({\bf Q})$ is large. And the magnetic form factor $f({\bf Q})$ can be 
looked up in tables~\cite{formfactor}.
Therefore, in order to determine the absolute magnitude of 
$\tilde{S}({\bf Q},\omega)$ from the measured intensity $\tilde{I}({\bf Q},E)$,
one needs to know the resolution volume $R_0$. 
In general, to obtain the resolution volume one could use the following 
references: (i) nuclear Bragg peaks;
(ii) sample incoherent elastic scattering; (iii) 
standard sample (e.g. vanadium) incoherent elastic scattering, and 
(iv) sample phonon scattering.  For single 
crystal samples, extinction in Bragg scattering can often
significantly affect the absolute Bragg intensities, and a good understanding 
of the instrument resolution is required, so (i) is less commonly
used nowadays. In this report, we will discuss some examples in our 
neutron scattering experiments where the magnetic scattering is normalized
using (ii) or (iv), providing formulas that one will be able to easily 
adapt to other measurements. We also  briefly discuss the case (iii).

\section{Experiment}

The experiments we discuss here include measurements on two different material
systems. The first system is a Fe-based superconductor FeTe$_{1-x}$Se$_x$, the
so-called ``11'' compound. Measurements on the ``11'' single crystal 
samples discussed
here were all performed on the BT7 and SPINS 
triple-axis-spectrometers at the NIST center for neutron research (NCNR)
(see Ref.~\onlinecite{Zhijun2010} for more details). 
We used horizontal beam collimations of
Guide-open-S-$80'$-$240'$ (S represents ``sample'') for the
inelastic scattering measurements on SPINS with fixed final energy
of 5 meV and a cooled Be filter after the sample to reduce
higher-order neutrons. At BT7, we used beam collimations of
open-$50'$-S-$50'$-$240'$ with fixed final energy of 14.7 meV and
two pyrolytic graphite filters after the sample.  The second system
is the multiferroic BiFeO$_3$. Neutron inelastic 
scattering experiments on the BiFeO$_3$ single crystal were performed on the 
ARCS time-of-flight
spectrometer at the Spallation Neutron Source (SNS) at Oak Ridge National 
laboratory (see Ref.~\onlinecite{BFO_Xu}). The data discussed here are those using a 40~meV neutron
incident energy. All the data are
described in reciprocal lattice units (r.l.u.) of $(a^*, b^*, c^*) =
(2\pi/a, 2\pi/b, 2\pi/c)$.

\section{Results and Discussion}

\subsection{Normalization with sample incoherent elastic scattering}

This is one of the most straight-forward methods in doing absolute
calibration. Since the cross-section for incoherent elastic 
scattering is quite simple:
\begin{equation}
\left.\frac{d\sigma}{d\Omega}\right\vert_{inc}^{el}=\frac{N}{4\pi}\sum_j \sigma^{inc}_je^{-2W_j},
\label{eqn:7}
\end{equation}
the summation is performed over all atoms in the unit cell, where $\sigma_j$ is 
the incoherent neutron scattering cross-section of the $j$th atom.
Based on Eq.~\ref{eqn:2}, the energy integrated incoherent elastic 
scattering intensity
\begin{equation}
\int \tilde{I}({\bf Q},E)dE=\frac{N}{4\pi}\sum_j \sigma^{inc}_je^{-2W_j}k_iR_0
\label{eqn:8}
\end{equation}
can be obtained by doing an energy scan through $\hbar\omega=0$ at a wave-vector
transfer ${\bf Q}$ far away from any magnetic/nuclear Bragg peaks.
For elastic scattering, $k_i$=$k_f$, therefore we have
\begin{equation}
Nk_fR_0=4\pi\frac{\int \tilde{I}({\bf Q},E)dE}
{\sum_j \sigma^{inc}_je^{-2W_j}}.
\label{eqn:9}
\end{equation}

\begin{table*}
\caption{Parameters for the two FeTe$_{1-x}$Se$_x$ samples. The integrated
intensities listed in the table are obtained from fits to the data sets
shown in Fig.~\ref{fig:1} (a) and (b), and have already been divided by 
monitor counts.}
\begin{ruledtabular}
\begin{tabular}{cccc}
  Sample &  $\int \tilde{I}({\bf Q},E)dE$ (meV)& $\sum_j \sigma^{inc}_j$ (barn) &$Nk_fR_0$ (meV$\cdot$barn$^{-1}$) \\

  \hline
    FeTe$_{0.7}$Se$_{0.3}$& $91.0/678000=1.34\times 10^{-4}$&$0.4+0.7\times 0.09+
0.3\times 0.32=0.559$ &$3.02\times 10^{-3}$\\
   FeTe$_{0.55}$Se$_{0.45}$ & $1128.8/140000= 8.03\times 10^{-3}$&
$0.4+0.55\times 0.09+ 0.45\times 0.32=0.594 $& $1.70\times 10^{-1}$ \\
\end{tabular}
\end{ruledtabular}
\label{tab:1}
\end{table*}

Now we use data from two samples as examples. 
The integrated incoherent elastic scattering intensities, the sum
of the incoherent scattering cross-sections, and the resolution volumes
are given in Table~\ref{tab:1}. If we neglect the variation of the 
Debye-Waller factor (assuming $e^{-2W} \sim 1$), knowing $Nk_fR_0$, 
we can then use 
Eq.~\ref{eqn:6} or Eq.~\ref{eqn:6a} to convert magnetic scattering 
into absolute units.

\begin{figure}[ht]
\includegraphics[trim=-1cm -2cm 0cm 0cm,clip=true,width=0.8\linewidth]{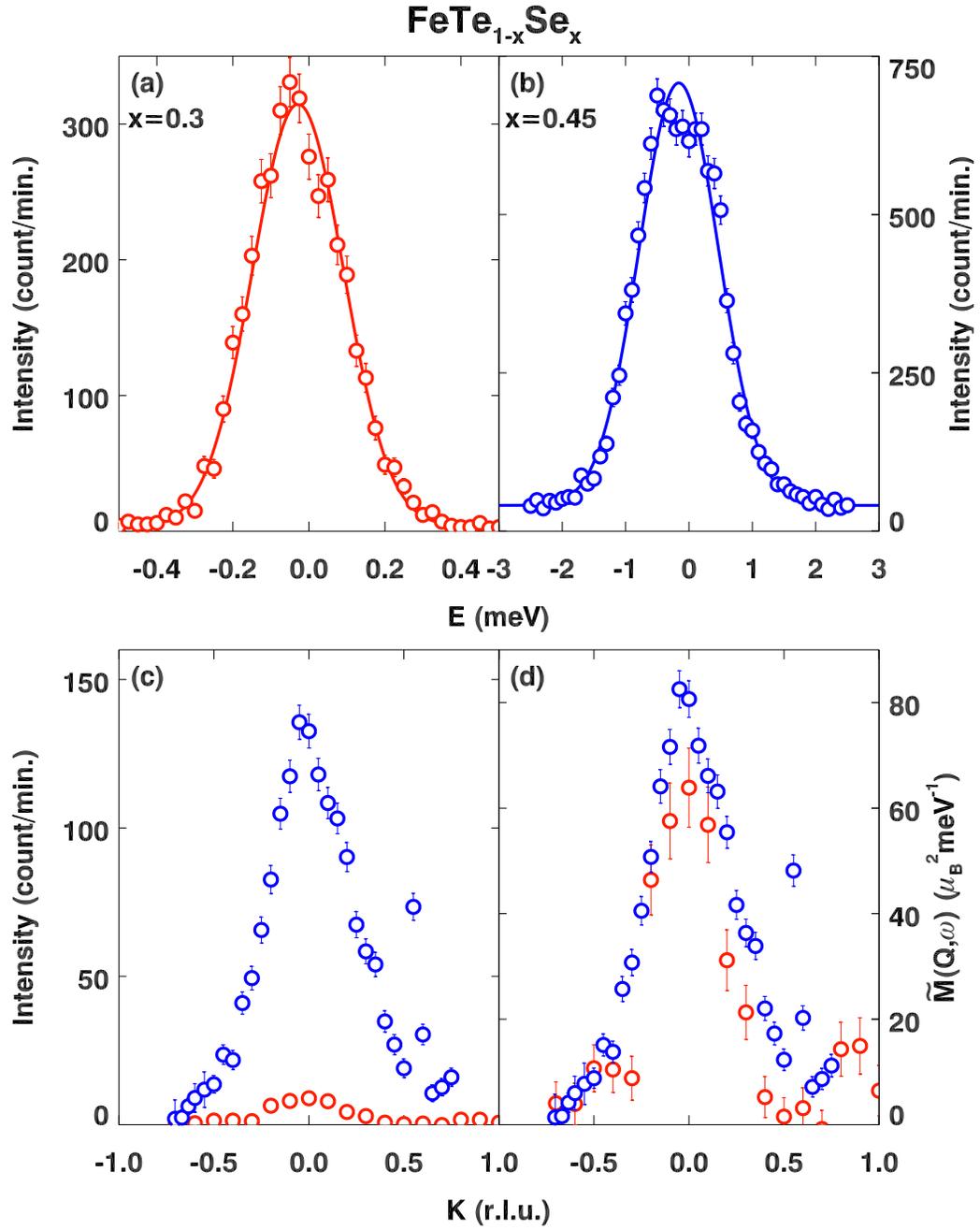}

\caption{(a) and (b): Incoherent elastic energy scans through 
$\hbar\omega=0$ taken from single crystals of FeTe$_{0.70}$Se$_{0.30}$ (red) 
and FeTe$_{0.55}$Se$_{0.45}$ (blue), 
measured at ${\bf Q}= (1.3,0.3,0)$, with monitor counts being 
$6.78 \times 10^5$ per minute,
and $1.4 \times 10^5$ per minute, performed on SPINS and BT7, respectively. 
(c): Constant energy
scans for magnetic scattering at $\hbar\omega=2$~meV, going along the 
transverse direction across ${\bf Q}=(0.5,0.5,0)$. The intensities
have been plotted as per minute. (d): Constant energy
scans for magnetic scattering at $\hbar\omega=2$~meV. The intensities
have been normalized to absolute units.
} \label{fig:1}
\end{figure}

Figures~\ref{fig:1} (c) and (d) show magnetic scattering measured at 
$\hbar\omega=2$~meV,
for ${\bf Q}$ along the [1$\bar{1}$0] direction across (0.5,0.5,0) 
from both samples 1 and 2. The measurements are performed on 
two similar samples with different weights, and on two different 
instruments (SPINS {\it vs.} BT7). Before the proper normalization, 
intensities per minute [Fig.~\ref{fig:1}~(c)] for the two measurements are 
obviously not comparable. Yet the results shown in units 
of $\mu_B^2\cdot$eV$^{-1}$
after normalization suggest that the two scans have very similar absolute 
intensities, as they should.

The procedure described above is one of our standard procedures for normalizing
magnetic scattering intensities using sample incoherent elastic scattering.
There are, of course, a few issues that need to be discussed.
First, the procedure described above was for triple-axis measurements 
using a fixed-E$_f$ 
mode, where $k_f$ would be a constant for all energy and wave-vector
transfers. If a constant-E$_i$ mode is used, $k_f$ would vary for different
energy transfers.
The volume $Nk_fR_0$ can still be obtained with the same methods. However, when 
using Eqs.\ref{eqn:6} or \ref{eqn:6a}, one will need to use the correct
$k_f$ for the measured energy transfer, not the $k_f$ for the incoherent
elastic scattering measurements (which is the same as $k_i$). The 
situation is simpler for measurements performed on time-of-flight instruments,
where the factors $k_i$ and $k_f$ have already been taken care of in 
properly reduced data sets available to users. One can use the same methods
to obtain the resolution volume $N\tilde{R}_0=Nk_fR_0$, which can later 
be used in Eqs.\ref{eqn:6} or \ref{eqn:6a} in place of $Nk_fR_0$, 
without having to worry about the change of $k_f$. 

Secondly, the number $N$ used here is the number of unit cells. Therefore 
the results correspond to squared moment/spin per formula unit. If there are 
more than one magnetic ions in a formula unit, the results need to be scaled
properly to obtain squared moment/spin per site if that is desired.

The advantage of using sample incoherent elastic scattering is of course its
simplicity. It is usually fast to perform an incoherent elastic scan, and no 
sample changing is required. However, for samples with relatively small 
incoherent scattering cross-sections, incoherent elastic scattering coming
from sample holders and sample environments can have a large contribution.
Therefore with this method it is common that one could over-estimate
the resolution volume, and under-estimate the real magnitude of magnetic 
scattering intensities.

\subsection{Normalization with standard sample 
incoherent elastic scattering}

Because vanadium has a large incoherent scattering cross-section, it
is often used as a standard sample for normalization purposes. The principle
is the same as described in the previous section. Nevertheless, in order
to use a vanadium sample for normalization, one needs to know the 
number of unit cells in both the measured sample and the vanadium
standard sample ($N$ in $Nk_fR_0$). A standard procedure would then be 
to carry out an incoherent elastic scan using the vanadium standard sample
under the same instrumental set-up; 
and use the energy integrated intensity and Eq.~\ref{eqn:9} to obtain 
$N_{vanadium}k_fR_0$. Knowing the weight of the vanadium sample and measured 
sample, one can then  obtain $N_{sample}k_fR_0$. 

The large incoherent scattering cross-section from vanadium makes incoherent
elastic background from the sample environment less of an issue in the 
normalization process, thus providing a significant advantage over 
the method using sample incoherent scattering. Standard vanadium
normalization runs can be readily performed without any detailed knowledge of 
the sample itself. This is another advantage of using vanadium normalization
over sample phonon normalization (see next subsection);
while for phonon normalization, one needs to have some basic
knowledge of the sample phonon spectra to know where in the Q-energy space
to perform the normalization runs. Vanadium normalization is the standard 
normalization method for a number of neutron facilities, including 
the ISIS neutron facility at U.K.

\subsection{Normalization with phonons}

Another common way of obtaining absolute calibration of the scattering intensity
without having to use a different standard sample, is to use the 
sample phonon scattering. The phonon scattering cross-section
can be written as:
\begin{equation}
\frac{d^2\sigma}{d\Omega dE} = N\frac{k_f}{k_i}S({\bf Q},\omega),
\label{eqn:phonon1}
\end{equation}
where $S({\bf Q},\omega)$ here is the response function. Typically
for normalization purposes, we measure phonons on the neutron energy 
loss side (i.e. $\hbar\omega > 0$). For acoustic phonons at small
${\bf q}={\bf Q}- {\bf G}$ values, $S({\bf Q},\omega)$ can be approximated to:
\begin{equation}
S({\bf Q},\omega)=\frac{n_q}{\omega({q})}|F_N({\bf G})|^2\frac{|{\bf Q \cdot
\xi}|^2}{2M}e^{-2W}\delta(\omega-\omega({q})).
\label{eqn:phonon2}
\end{equation}
Here ${\bf G}$ is the Bragg wave-vector near which the phonon is being measured.
$n_q=\frac{1}{1-e^{-\hbar\omega/k_BT}}$ is the Bose factor, $F_N({\bf G})$
is the acoustic phonon structure factor which is the same as the Bragg structure
factor at ${\bf G}$. $\xi$ is the unit vector along the phonon polarization 
direction, and $|{\bf Q \cdot \xi}|^2$ gives the ``polarization factor''. 
$M=\sum_j m_j$ is the summation of the atomic mass in the unit cell.

Phonon measurements can typically be performed as constant-${\bf Q}$ or 
constant-E scans. For a constant-${\bf Q}$ scan, the energy integrated 
phonon intensity can be written as:
\begin{equation}
\int \tilde{I}({\bf Q},E)dE=\frac{n_q}{\hbar\omega({q})}\frac{(\hbar{\bf Q})^2}{2m}
\frac{m}{M}\cdot \cos^2\beta |F_N({\bf G})|^2e^{-2W}Nk_fR_0.
\label{eqn:phonon3}
\end{equation}
Here $m$ is the mass of neutron. Rewriting $S({\bf Q},\omega)$ in this 
format makes it easier to put numbers in. One will not have to write 
atomic weights in units of kg or g, but can rather use the inverse atomic
numbers in place of $\frac{m}{M}$.  $\frac{(\hbar{\bf Q})^2}{2m}$ is
then the neutron energy at wave-vector ${\bf Q}$ which is practically just
$2.0717{\bf Q}^2$ in units of meV where ${\bf Q}$ should have units of
\AA$^{-1}$. $\beta$ is the angle between ${\bf Q}$ and $\xi$.

Another approach is to perform constant-E scans where the {\bf Q}-integrated
phonon energy can be obtained:
\begin{equation}
\int \tilde{I}({\bf Q},E)d{\bf q}=\frac{1}{d\omega/dq}\frac{n_q}{\hbar\omega(q)}\frac{(\hbar{\bf Q})^2}{2m}
\frac{m}{M}\cdot \cos^2\beta |F_N({\bf G})|^2e^{-2W}Nk_fR_0.
\label{eqn:phonon4}
\end{equation}
Here $d\omega/dq$ is the phonon velocity at the measured energy/wave-vector. 
The unit used for $q$ should be the same in the intensity integral and 
phonon velocity on both sides of the equation so that they can cancel out.

We use examples from our measurements of BiFeO$_3$ single crystals on the 
time-of-flight spectrometer ARCS at SNS as an example to show how one
can obtain the resolution volume $N\tilde{R}_0$ with phonons. In 
Table~\ref{tab:2}, the numbers are given for longitudinal phonons measured
near (110), (300), and (100) Bragg peaks. We also assume that the Debye-Waller factor
$e^{-2W}$ is close to 1. As discussed previously, the time-of-flight data
have already been treated so that the factors $k_i$ and $k_f$ are removed,
and we use $N\tilde{R}_0$ instead of  $Nk_fR_0$.

\begin{table*}
\caption{Parameters for longitudinal 
phonon measurements on BiFeO$_3$. The first two rows
are results  from constant-E cuts near ${\bf G}=$(3,0,0) and (1,1,0) [see
Fig.~\ref{fig:2} (a) and (b)]; 
and the last row are results from a constant-${\bf Q}$ cut [see Fig.~\ref{fig:2} (c)]
at ${\bf Q}=$(1.3,0,0).  
For all measurements, $T=300$~K, $\frac{m}{M}=\frac{1}{313}$, $\cos^2\beta=1.0$ . }
\begin{ruledtabular}
\begin{tabular}{cccccccc}
  ${\bf Q}$ (r.l.u.) &$ \hbar\omega$ (meV) & $d\omega/dq$ (meV$\cdot$\AA$^{-1}$)
  & $n_q$ &$\frac{\hbar{\bf Q}^2}{2m}$ (meV) 
& $\int \tilde{I}({\bf Q},E)dq$ (\AA$^{-1}$)& $|F_N({\bf G})|^2$ (barn) 
&$N\tilde{R}_0$ (meV$\cdot$barn$^{-1}$) \\
  \hline
(3,0,0) & $5.0$ & $20.85$ & $5.69$ & $46.95$ & $0.00198$ &
$868.2$ & $2.78\times10^{-4}$\\
(1,1,0) & $4.0$ & $18.37$ & $6.98$ & $10.43$ & $0.00124$ &
$1653.31$ & $2.37\times10^{-4}$\\
\hline
  & & & & & $\int \tilde{I}({\bf Q}, E)dE (meV^{-1})$ & & \\
  \hline
(1.3,0,0) & $6.92$ &  & $4.26$ & $8.82$ & $0.00338$ &
$651.1$ & $3.0\times10^{-4}$\\
\hline
   
\end{tabular}
\end{ruledtabular}
\label{tab:2}
\end{table*}

\begin{figure}[ht]
\includegraphics[trim=0cm -2cm 0cm -1.5cm,clip=true,width=0.8\linewidth]{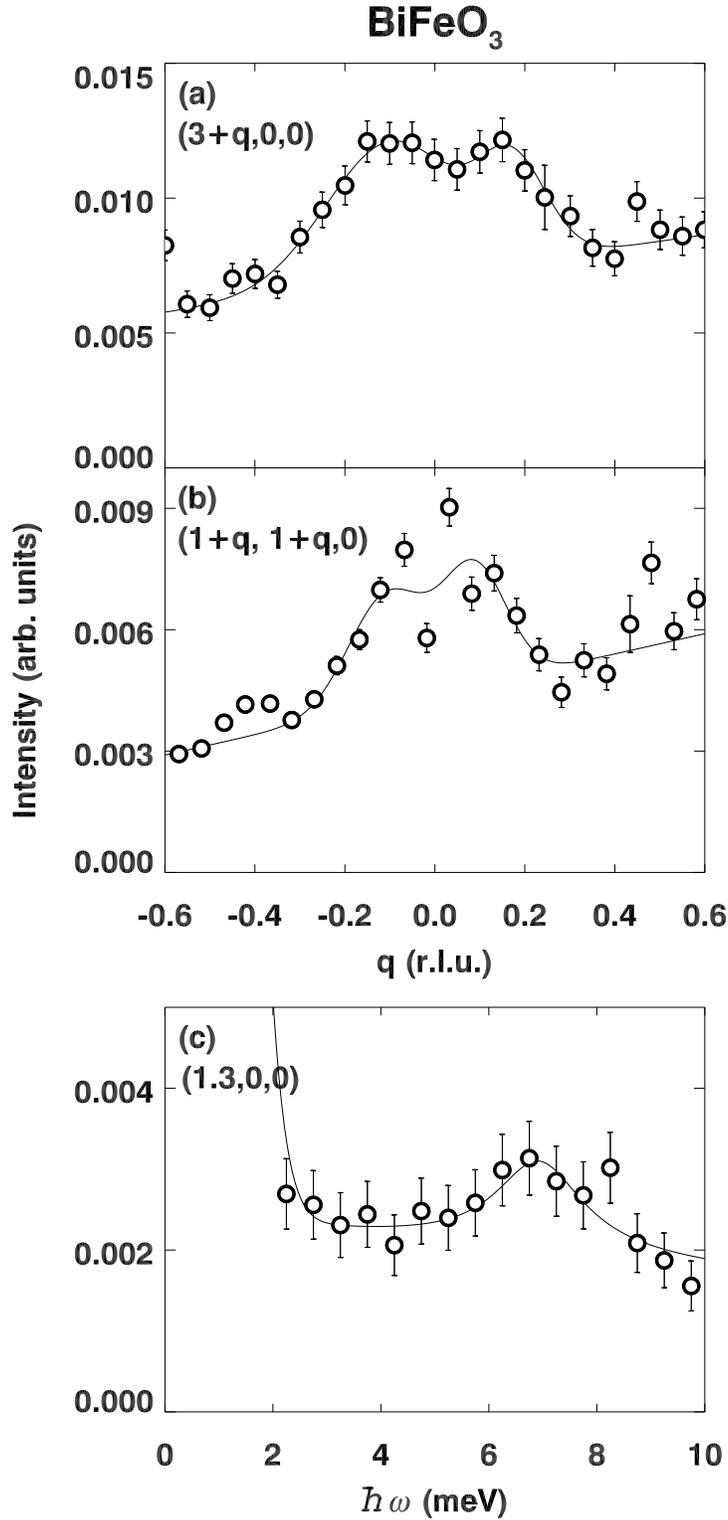}

\caption{(a) and (b): Constant-Energy cuts through $\hbar\omega=5$ and 
$4$~meV near ${\bf Q}=(3,0,0)$ and $(1,1,0)$ along the longitudinal direction. 
Here the phonon intensities include phonons on both sides (short-${\bf Q}$
and long-${\bf Q}$) of the Bragg peaks, and are fitted with double-Gaussians.
The $q$-integrated intensities listed in Table~\ref{tab:2} use half of the 
double-Gaussian areas. The sample is a BiFeO$_3$ single crystal, measured on 
ARCS at SNS. 
(c) A constant-${\bf Q}$ cut taken at ${\bf Q}=(1.3,0,0)$.} 
\label{fig:2}
\end{figure}

The resolution volume $NR_0$ from the three phonon profiles 
shown in Table~\ref{tab:2} and Fig.~\ref{fig:2} 
are consistent within expected error bars.
After obtaining $N\tilde{R}_0$, we can use Eqs.~\ref{eqn:6} and \ref{eqn:6a}
to normalize magnetic scattering intensities~\cite{BFO_Xu}. 
Assuming $g=2$, we were able to calculate the integrated spectral
weight and a spin per site of $S \sim 2.1$ per Fe, which is in good 
agreement with the theoretical expectation of $S=5/2$ for Fe$^{3+}$.

The advantage of using phonon scattering for normalization is that by avoiding
the use of elastic scattering, there is less background 
contribution in the normalization process. Nevertheless, other factors such as
phonon aharmonicity, variation of the Debye-Waller factor and uncertainties
in the structure factor $|F_N({\bf G})|^2$ can still lead to systematic errors.
It is reasonable to expect a systematic error in the range of $\sim 20\%$ in
these normalization processes.

\section{Summary}

In this paper, we have discussed 
various methods for performing absolute normalization
of magnetic inelastic neutron scattering data. The resolution volume can 
be obtained  using Eqs.~\ref{eqn:9}, \ref{eqn:phonon3}, or \ref{eqn:phonon4},
depending on the reference chosen for the normalization. Data-sets
from previous experiments are given as examples for the normalization process.
The modified dynamic spin correlation function $\tilde{S}({\bf Q},\omega)$,
or the imaginary part of the dynamic susceptibility ${\chi}''({\bf Q},\omega)$,
have been clearly defined in the paper; and these quantities 
can then be obtained using Eq.~\ref{eqn:6}, \ref{eqn:6a}, or \ref{eqn:6c}. 
The formulas described here are for general purposes and independent
of instrument configurations. We hope they can serve as easy-to-use
references for future inelastic neutron scattering measurements.

\section{Acknowledgments}
We would like to thank S. M. Shapiro, and I. Zaliznyak for 
useful discussions. Technical support from instrument scientists M. Stone, 
D. Singh, and Y. Zhao is also gratefully acknowledged. Financial support by 
Office of Basic Energy Sciences, U.S. Department of Energy 
under contract No. DE-AC02-98CH10886 is acknowledged.


\begin{thebibliography}{9}%
\makeatletter
\providecommand \@ifxundefined [1]{%
 \@ifx{#1\undefined}
}%
\providecommand \@ifnum [1]{%
 \ifnum #1\expandafter \@firstoftwo
 \else \expandafter \@secondoftwo
 \fi
}%
\providecommand \@ifx [1]{%
 \ifx #1\expandafter \@firstoftwo
 \else \expandafter \@secondoftwo
 \fi
}%
\providecommand \natexlab [1]{#1}%
\providecommand \enquote  [1]{``#1''}%
\providecommand \bibnamefont  [1]{#1}%
\providecommand \bibfnamefont [1]{#1}%
\providecommand \citenamefont [1]{#1}%
\providecommand \href@noop [0]{\@secondoftwo}%
\providecommand \href [0]{\begingroup \@sanitize@url \@href}%
\providecommand \@href[1]{\@@startlink{#1}\@@href}%
\providecommand \@@href[1]{\endgroup#1\@@endlink}%
\providecommand \@sanitize@url [0]{\catcode `\\12\catcode `\$12\catcode
  `\&12\catcode `\#12\catcode `\^12\catcode `\_12\catcode `\%12\relax}%
\providecommand \@@startlink[1]{}%
\providecommand \@@endlink[0]{}%
\providecommand \url  [0]{\begingroup\@sanitize@url \@url }%
\providecommand \@url [1]{\endgroup\@href {#1}{\urlprefix }}%
\providecommand \urlprefix  [0]{URL }%
\providecommand \Eprint [0]{\href }%
\providecommand \doibase [0]{http://dx.doi.org/}%
\providecommand \selectlanguage [0]{\@gobble}%
\providecommand \bibinfo  [0]{\@secondoftwo}%
\providecommand \bibfield  [0]{\@secondoftwo}%
\providecommand \translation [1]{[#1]}%
\providecommand \BibitemOpen [0]{}%
\providecommand \bibitemStop [0]{}%
\providecommand \bibitemNoStop [0]{.\EOS\space}%
\providecommand \EOS [0]{\spacefactor3000\relax}%
\providecommand \BibitemShut  [1]{\csname bibitem#1\endcsname}%
\let\auto@bib@innerbib\@empty
\bibitem [{\citenamefont {Broholm}(1989)}]{Broholm1}%
  \BibitemOpen
  \bibfield  {author} {\bibinfo {author} {\bibfnamefont {C.~L.}\ \bibnamefont
  {Broholm}},\ }\emph {\bibinfo {title} {Magnetic Fluctuations in Heavy Fermion
  Systems}},\ \href@noop {} {Ph.D. thesis},\ \bibinfo  {school} {Riso National
  Laboratory} (\bibinfo {year} {1989})\BibitemShut {NoStop}%
\bibitem [{\citenamefont {Steinsvoll}\ \emph {et~al.}(1984)\citenamefont
  {Steinsvoll}, \citenamefont {Majkrzak}, \citenamefont {Shirane},\ and\
  \citenamefont {Wicksted}}]{Steinsvoll}%
  \BibitemOpen
  \bibfield  {author} {\bibinfo {author} {\bibfnamefont {O.}~\bibnamefont
  {Steinsvoll}}, \bibinfo {author} {\bibfnamefont {C.~F.}\ \bibnamefont
  {Majkrzak}}, \bibinfo {author} {\bibfnamefont {G.}~\bibnamefont {Shirane}}, \
  and\ \bibinfo {author} {\bibfnamefont {J.}~\bibnamefont {Wicksted}},\
  }\href@noop {} {\bibfield  {journal} {\bibinfo  {journal} {Physical Review
  B}\ }\textbf {\bibinfo {volume} {30}},\ \bibinfo {pages} {2377} (\bibinfo
  {year} {1984})}\BibitemShut {NoStop}%
\bibitem [{\citenamefont {Ziebeck}\ and\ \citenamefont
  {Brown}(1980)}]{Ziebeck}%
  \BibitemOpen
  \bibfield  {author} {\bibinfo {author} {\bibfnamefont {K.~R.~A.}\
  \bibnamefont {Ziebeck}}\ and\ \bibinfo {author} {\bibfnamefont {P.~J.}\
  \bibnamefont {Brown}},\ }\href@noop {} {\bibfield  {journal} {\bibinfo
  {journal} {Journal of Physics F-Metal Physics}\ }\textbf {\bibinfo {volume}
  {10}},\ \bibinfo {pages} {2015} (\bibinfo {year} {1980})}\BibitemShut
  {NoStop}%
\bibitem [{\citenamefont {Chesser}\ and\ \citenamefont {Axe}(1973)}]{Chesser1}%
  \BibitemOpen
  \bibfield  {author} {\bibinfo {author} {\bibfnamefont {N.~J.}\ \bibnamefont
  {Chesser}}\ and\ \bibinfo {author} {\bibfnamefont {J.~D.}\ \bibnamefont
  {Axe}},\ }\href@noop {} {\bibfield  {journal} {\bibinfo  {journal} {Acta
  Crystallographica Section A}\ }\textbf {\bibinfo {volume} {A 29}},\ \bibinfo
  {pages} {160} (\bibinfo {year} {1973})}\BibitemShut {NoStop}%
\bibitem [{\citenamefont {Shirane}\ \emph {et~al.}(2002)\citenamefont
  {Shirane}, \citenamefont {Shapiro},\ and\ \citenamefont
  {Tranquada}}]{Shiranebook}%
  \BibitemOpen
  \bibfield  {author} {\bibinfo {author} {\bibfnamefont {G.}~\bibnamefont
  {Shirane}}, \bibinfo {author} {\bibfnamefont {S.~M.}\ \bibnamefont
  {Shapiro}}, \ and\ \bibinfo {author} {\bibfnamefont {J.}~\bibnamefont
  {Tranquada}},\ }\href@noop {} {\emph {\bibinfo {title} {Neutron Scattering
  with a Triple-Axis Spectrometer Basic Techniques}}}\ (\bibinfo  {publisher}
  {Cambridge University Press},\ \bibinfo {year} {2002})\BibitemShut {NoStop}%
\bibitem [{\citenamefont {Marshall}\ and\ \citenamefont
  {Lowde}(1968)}]{Marshall}%
  \BibitemOpen
  \bibfield  {author} {\bibinfo {author} {\bibfnamefont {W.}~\bibnamefont
  {Marshall}}\ and\ \bibinfo {author} {\bibfnamefont {R.~D.}\ \bibnamefont
  {Lowde}},\ }\href@noop {} {\bibfield  {journal} {\bibinfo  {journal} {Rep.
  Prog. Phys.}\ }\textbf {\bibinfo {volume} {31}},\ \bibinfo {pages} {705}
  (\bibinfo {year} {1968})}\BibitemShut {NoStop}%
\bibitem [{\citenamefont {Brown}()}]{formfactor}%
  \BibitemOpen
  \bibfield  {author} {\bibinfo {author} {\bibfnamefont {P.~J.}\ \bibnamefont
  {Brown}},\ }\enquote {\bibinfo {title} {International tables for
  crystallography},}\ Chap.\ \bibinfo {chapter} {4.4.5 Magnetic form factors},
  pp.\ \bibinfo {pages} {391--399}\BibitemShut {NoStop}%
\bibitem [{\citenamefont {Xu}\ \emph {et~al.}(2010)\citenamefont {Xu},
  \citenamefont {Wen}, \citenamefont {Xu}, \citenamefont {Jie}, \citenamefont
  {Lin}, \citenamefont {Li}, \citenamefont {Chi}, \citenamefont {Singh},
  \citenamefont {Gu},\ and\ \citenamefont {Tranquada}}]{Zhijun2010}%
  \BibitemOpen
  \bibfield  {author} {\bibinfo {author} {\bibfnamefont {Z.}~\bibnamefont
  {Xu}}, \bibinfo {author} {\bibfnamefont {J.}~\bibnamefont {Wen}}, \bibinfo
  {author} {\bibfnamefont {G.}~\bibnamefont {Xu}}, \bibinfo {author}
  {\bibfnamefont {Q.}~\bibnamefont {Jie}}, \bibinfo {author} {\bibfnamefont
  {Z.}~\bibnamefont {Lin}}, \bibinfo {author} {\bibfnamefont {Q.}~\bibnamefont
  {Li}}, \bibinfo {author} {\bibfnamefont {S.}~\bibnamefont {Chi}}, \bibinfo
  {author} {\bibfnamefont {D.~K.}\ \bibnamefont {Singh}}, \bibinfo {author}
  {\bibfnamefont {G.}~\bibnamefont {Gu}}, \ and\ \bibinfo {author}
  {\bibfnamefont {J.~M.}\ \bibnamefont {Tranquada}},\ }\href@noop {} {\bibfield
   {journal} {\bibinfo  {journal} {Phys. Rev. B}\ }\textbf {\bibinfo {volume}
  {82}},\ \bibinfo {pages} {104525} (\bibinfo {year} {2010})}\BibitemShut
  {NoStop}%
\bibitem [{\citenamefont {Xu}\ \emph {et~al.}(2012)\citenamefont {Xu},
  \citenamefont {Wen}, \citenamefont {Berlijn}, \citenamefont {Gehring},
  \citenamefont {Stock}, \citenamefont {Stone}, \citenamefont {Ku},
  \citenamefont {Gu}, \citenamefont {Shapiro}, \citenamefont {Birgeneau},\ and\
  \citenamefont {Xu}}]{BFO_Xu}%
  \BibitemOpen
  \bibfield  {author} {\bibinfo {author} {\bibfnamefont {Z.}~\bibnamefont
  {Xu}}, \bibinfo {author} {\bibfnamefont {J.}~\bibnamefont {Wen}}, \bibinfo
  {author} {\bibfnamefont {T.}~\bibnamefont {Berlijn}}, \bibinfo {author}
  {\bibfnamefont {P.~M.}\ \bibnamefont {Gehring}}, \bibinfo {author}
  {\bibfnamefont {C.}~\bibnamefont {Stock}}, \bibinfo {author} {\bibfnamefont
  {M.~B.}\ \bibnamefont {Stone}}, \bibinfo {author} {\bibfnamefont
  {W.}~\bibnamefont {Ku}}, \bibinfo {author} {\bibfnamefont {G.}~\bibnamefont
  {Gu}}, \bibinfo {author} {\bibfnamefont {S.~M.}\ \bibnamefont {Shapiro}},
  \bibinfo {author} {\bibfnamefont {R.~J.}\ \bibnamefont {Birgeneau}}, \ and\
  \bibinfo {author} {\bibfnamefont {G.}~\bibnamefont {Xu}},\ }\href@noop {}
  {\bibfield  {journal} {\bibinfo  {journal} {Physical Review B}\ }\textbf
  {\bibinfo {volume} {86}},\ \bibinfo {pages} {174419} (\bibinfo {year}
  {2012})}\BibitemShut {NoStop}%
\end{thebibliography}
%

\end{document}